\documentclass[prb,11pt,nofootinbib,superscriptaddress,longbibliography]{revtex4-2}

\usepackage{graphicx}
\usepackage{hyperref}
\usepackage{amsmath}
\usepackage{amssymb}
\usepackage[usenames,dvipsnames]{color}

\usepackage{xcolor}
\usepackage{soul}
\soulregister\cite7
\soulregister\ref7
\soulregister\pageref7

\begin{document}

\title{Lattice Reconstruction and Orbital Hybridization Suppress Magnetism in TaCo$_2$Te$_2$}
%\title{Intertwined Peierls Distortions and Canted Magnetism in the layered Material TaCo$_2$Te$_2$ }

\date{\today}

\author{Ulysse Chazarin $^{\dagger}$}
\altaffiliation{These authors contributed equally to this work.}
\affiliation{Department of Physics, Nanoscience Center, 
University of Jyväskyl\"a, FI-40014 University of Jyväskyl\"a, Finland}
\author{Bharat C. Bathu}
\altaffiliation{These authors contributed equally to this work.}
\affiliation{Department of Physics, Nanoscience Center, 
University of Jyväskyl\"a, FI-40014 University of Jyväskyl\"a, Finland}
\altaffiliation{These authors contributed equally to this work.}
\author{Zuned Ahmed}
\affiliation{Department of Physics, Nanoscience Center, 
University of Jyväskyl\"a, FI-40014 University of Jyväskyl\"a, Finland}
\author{Marta Zonno}
\affiliation{Synchrotron SOLEIL, L’Orme des Merisiers, Saint-Aubin, France}
\author{Chiara Bigi}
\affiliation{Synchrotron SOLEIL, L’Orme des Merisiers, Saint-Aubin, France}
\author{Francois Bertran}
\affiliation{Synchrotron SOLEIL, L’Orme des Merisiers, Saint-Aubin, France}
\author{Adolfo O. Fumega}
\affiliation{Department of Applied Physics, Aalto University, Espoo, Finland}
\author{Orlando J. Silveira $^{\dagger}$}
\affiliation{Department of Applied Physics, Aalto University, Espoo, Finland}
\author{Shawulienu Kezilebieke}
\email{Corresponding author. Email: kezilebieke.a.shawulienu@jyu.fi, ulysse.u.chazarin@jyu.fi, orlando.silveirajunior@aalto.fi}
\affiliation{Department of Physics, Nanoscience Center, 
University of Jyväskyl\"a, FI-40014 University of Jyväskyl\"a, Finland}
\affiliation{ Department of Chemistry, Nanoscience Center, 
University of Jyväskyl\"a, FI-40014 University of Jyväskyl\"a, Finland}

\keywords{Scanning tunneling microscopy and spectroscopy,  Van der Waals material, Magnetism, orbital selective electron reconstruction}

\begin{abstract}
Structural reconstruction in low-dimensional quantum materials can strongly modify electronic symmetry and magnetic stability through orbital hybridization. Here, we investigate the interplay between lattice reconstruction, electronic structure, and magnetic instability in the layered van der Waals compound TaCo$_2$Te$_2$ using scanning tunneling microscopy and spectroscopy (STM/STS), non-contact atomic force microscopy (nc-AFM), angle-resolved photoemission spectroscopy (ARPES), and density functional theory (DFT). While nc-AFM resolves a distorted hexagonal Te surface lattice, STM/STS reveal a pronounced square-like electronic symmetry that does not directly follow the atomic structure. ARPES further shows a strongly anisotropic Fermi surface and reconstructed low-energy states. Spatially resolved spectroscopy and orbital-projected DFT demonstrate that the bias-dependent STM contrast does not arise from a simple reversal between occupied and unoccupied states, but from the energy-integrated local density of states dominated by electronic states exhibiting opposite spatial contrast at selected energies. DFT calculations further show that reconstruction suppresses the magnetic instability present in the undistorted structure, stabilizing a nonmagnetic ground state through enhanced orbital hybridization. These results establish TaCo$_2$Te$_2$ as a model system in which lattice reconstruction reorganizes electronic symmetry and suppresses magnetism, highlighting structural reconstruction as a route for controlling correlated and magnetic phases in low-dimensional quantum materials.

\end{abstract}
\maketitle

\newpage

\section*{Introduction}

Low-dimensional magnetic materials provide a powerful platform for engineering emergent electronic phases relevant for spintronic and quantum technologies.\cite{Huang2017-fx,Gong2017-kn, Burch2018-ii,Gibertini2019-ug}.
In particular, spintronic devices rely on the precise control of spin at reduced length scales, motivating intense efforts to understand and manipulate magnetism in atomically thin systems\cite{mi_two-dimensional_2023,zhao_novel_2025}. The emergence of layered van der Waals (vdW) magnets has enabled unprecedented control over magnetic order through exfoliation down to the monolayer limit and subsequent integration into heterostructures, twisted systems, and nanoscale devices\cite{algarni_recent_2025,witte_tuning_2024,cheon_nature_2025,Jiang2018-ig,Song2018-ai,Wang2023-zj}.

Several families of vdW magnetic materials have now been identified, including chromium halides and intercalated transition-metal chalcogenides\cite{Huang2017-fx,zhang_direct_2019,iturriaga_magnetic_2023,wang_interfacial_2023}. While chromium-based compounds such as CrI$_3$ and CrBr$_3$ represent prototypical insulating ferromagnets that retain magnetic order in the two-dimensional limit and can be realized using techniques such as molecular beam epitaxy\cite{ Li2020-vz,Chen2019-yb,Kezilebieke2020-de,Kezilebieke2021-hw}. More recently, intercalated dichalcogenides including  Fe$_3$GeTe$_2$ and Fe$_4$GeTe$_2$ have attracted increasing attention because of their enhanced air stability, tunable electronic properties, and compatibility with conventional growth techniques\cite{xie_air_2024,iturriaga_magnetic_2023,wang_interfacial_2023}. 

Within this broader family, TaCo$_2$Te$_2$ provides a particularly intriguing platform because it combines localized Co moments with pronounced structural instabilities and unconventional electronic properties while retaining the environmental stability characteristic of layered chalcogenides. Previous studies have reported Dirac-like topological features, Peierls distortion of the intercalated Co atoms as well as signatures of high-temperature structural modulations \cite{rong_realization_2023,jiao_large_2025,mathur_emergence_2024}. The magnetic ground state, however, remains unresolved: experimental reports have variously suggested canted magnetic behavior\cite{singha_taco2te2_2022} and unusual field-dependent responses,\cite{wang_effect_2022,pradhan_magnetic_2026} while other work has proposed a nonmagnetic ground state\cite{wang_effect_2022}. These apparently contradictory observations raise the possibility that an intrinsic lattice reconstruction plays a central role in reshaping the low-energy electronic structure and suppressing the expected magnetic state.

Reconstruction-driven electronic instabilities are increasingly recognized as mechanisms capable of destabilizing magnetism in low-dimensional systems. In monolayer VSe$_2$, for example, charge-density-wave formation suppresses the expected ferromagnetic phase through long-range electronic reconstruction \cite{bonilla_strong_2018,kezilebieke_electronic_2020,fumega_absence_2019}. Strong spin fluctuations similarly prevent long-range magnetic order in FeSe\cite{shishidou_magnetic_2018}. In other systems, orbital hybridization can suppress magnetism more directly by modifying exchange interactions and redistributing spectral weight across competing orbitals, as demonstrated in Cr$_2$Te$_{1-x}$W$_x$O$_6$ \cite{zhu_tuning_2014}. Whether analogous mechanisms operate in TaCo$_2$Te$_2$ requires direct, atomic-scale access to the interplay between lattice distortion, electronic reconstruction, and orbital-selective magnetism.

%More broadly, reconstruction-driven electronic instabilities are increasingly recognized as mechanisms capable of destabilizing magnetism in low-dimensional systems. For example, charge-density-wave formation in monolayer VSe$_2$ suppresses the expected ferromagnetic phase through long-range electronic reconstruction\cite{bonilla_strong_2018,kezilebieke_electronic_2020}. Similarly, strong spin fluctuations prevent the establishment of long-range magnetic order in FeSe \cite{shishidou_magnetic_2018}. In other systems, orbital hybridization itself can suppress magnetism by modifying exchange interactions and redistributing spectral weight across competing orbitals, as demonstrated in Cr$_2$Te$_{1-x}$W$_x$O$_6$ \cite{zhu_tuning_2014}. Determining whether similar mechanisms operate in TaCo$_2$Te$_2$ therefore, requires direct access to the interplay between lattice distortion, electronic reconstruction, and orbital-selective magnetism at the atomic scale.

Here, we combine scanning tunneling microscopy and spectroscopy (STM/STS), non-contact atomic force microscopy (nc-AFM), angle-resolved photoemission spectroscopy (ARPES), and density functional theory (DFT) calculations to demonstrate that the surface distortion reorganizes the electronic symmetry of TaCo$_2$Te$_2$. The reorganization is observed as a contrast modulation in the STM topography, which does not originate from a simple reversal between occupied and unoccupied states, but rather from the energy-integrated LDOS dominated by specific electronic states exhibiting opposite spatial contrast at selected energies. We show that this distortion-enhanced hybridization reconstructs the low-energy electronic structure and suppresses magnetic ordering through orbital-selective electronic reconstruction.

\section*{Results and discussion}

\subsection*{Structural properties}

\begin{figure}[h!]
\begin{center}
\includegraphics[width=\columnwidth]{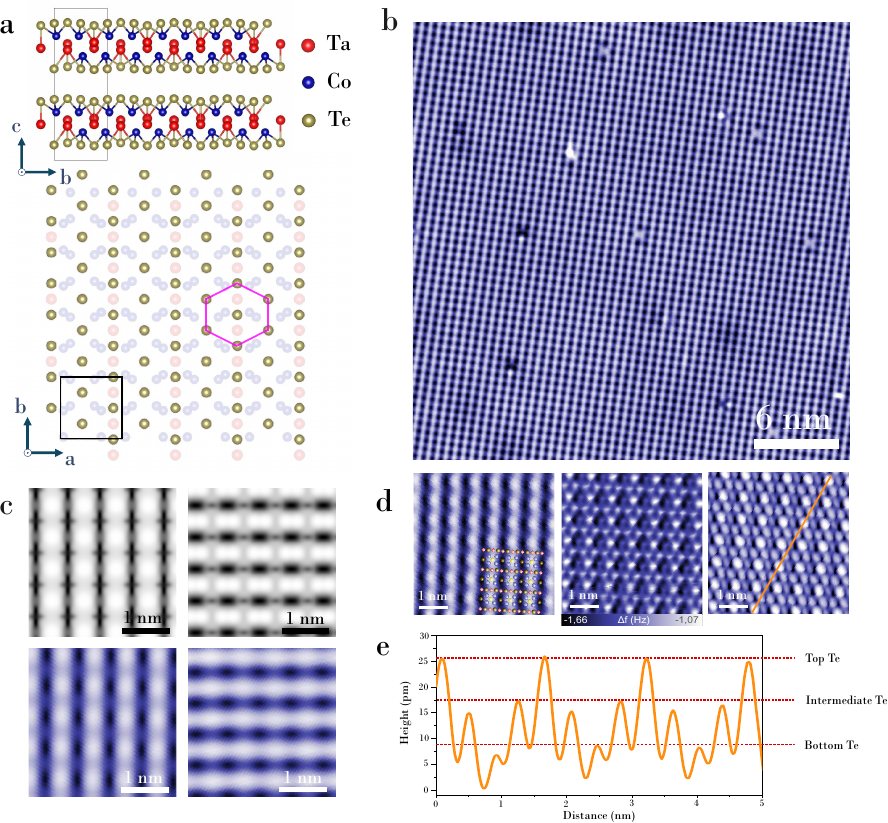}
\caption{(a) Top panel : Structural model of the TaCo$_2$Te$_2$ lattice including Peierl's distortion of the Co atoms. The unit cell is indicated by the black rectangle. Bottom Panel : Top view of Te surface lattice, with Co and Ta atomic positions for reference. A hexagonal motif is highlighted in purple. (b) Constant-current STM image of the TaCo$_2$Te$_2$ surface (V = 1V, I = 300pA, $30\times30$ nm$^2$). (c) Comparison between experimental (bottom: V = -200mV (left) / 200mV (right), I = 300pA, $3\times3$ nm$^2$ ) and simulated (top: V = -100mV (left) / 100meV (right)) constant current STM images. (d) Atomic resolution images of TaCo$_2$Te$_2$. Left : Constant-current STM topography (-400mV, 300pA, $5\times5$ nm$^2$), Middle : Constant-height nc-AFM $\Delta f$ image (z=-220pm relative to set point (-200mV, 100pA), A = 800pm, $5\times5$ nm$^2$) and right : Constant-$\Delta f$ nc-AFM topography (0V, $\Delta f$ = - 4.5 Hz, A = 1nm, $5\times5$ nm$^2$ ). A schematic of the atomic lattice is overlaid on the STM image. (e) Line profile taken along an atomic row (yellow line in (d), right panel). The red dashed lines highlight the height difference between surface Te atoms.}
\label{Fig.1}
\end{center}
\end{figure}

TaCo$_2$Te$_2$ is a layered van der Waals (vdW) material composed of sextuple atomic layers, as illustrated in Figure~\ref{Fig.1}(a). Along its crystallographic \textit{c-}direction, both the top and bottom layers of each TaCo$_2$Te$_2$ unit are terminated by Te atoms, with one Te layer becoming the exposed surface after cleavage. In the ideal structure, the topmost Te layer is arranged in a hexagonal lattice. However, our DFT structural relaxations together with constant-current STM measurements of the TaCo$_2$Te$_2$ surface reveal pronounced structural and electronic distortions that prevent direct visualization of the complete Te lattice in STM images. Instead, the surface electronic states exhibit a square lattice pattern characterized by bright protrusions separated by darker regions along one crystallographic direction and enhanced intensity along the orthogonal bonds (Figure~\ref{Fig.1}(b)). We measure averaged lattice parameters of the square network of $a_{STM} = b_{STM} = 0.67\  \text{nm}$. As the energy approaches the Fermi level ($E_F$ = 0) , this anisotropy becomes increasingly pronounced along the brighter bond direction, consistent with the reported distortion in \cite{mathur_emergence_2024}. Notably, the square electronic pattern rotates by 90$^\circ$ upon reversal of the sample bias, as shown in Figure~\ref{Fig.1}(c), indicating a strong energy dependence of the reconstructed local density of states (more STM images are displayed in Fig.SI-1(a-i)). Structural relaxation of both bulk and monolayer TaCo$_2$Te$_2$ with DFT generates inequivalent Te sublattices with reduced local symmetry (square symmetry instead of hexagonal), and STM simulations of the monolayer TaCo$_2$Te$_2$ shown in Figure \ref{Fig.1}(c) display the same bias dependent patterns as in the experiment.
%Our simulations suggest that the square-like electronic patterns come from the slightly elevated atomics position of the Te atoms, although the reason for the pattern  with the bias is not clear considering the STM images alone. 

Since STM topography contains both geometric and local density of states contributions, we performed non-contact Atomic Force Microscopy (nc-AFM) measurements to disentangle the electronic contribution and experimentally resolve the positions of the surrounding Te atoms, as shown in Fig.~\ref{Fig.1}. The nc-AFM image acquired in constant-height $\Delta f$ mode (Fig.~\ref{Fig.1}(d), middle panel) reveals the hexagonal arrangement of the top Te layer, consistent with previous AFM studies on TaCo$_2$Te$_2$ \cite{neuhausen_scanning_1998}. From the nc-AFM data, we extract in-plane hexagonal lattice parameters of $a_{AFM} = b_{AFM} = 0.37\ \text{nm}$. 

Finally, constant-frequency-shift nc-AFM topography (Fig.~\ref{Fig.1}(d), right panel) was performed to evaluate the relative height variations of the surface Te atoms. A line profile extracted along a crystallographic direction is presented in Fig.~\ref{Fig.1}(e). Within a single unit cell, three distinct groups of atoms can be identified, corresponding to one top, two intermediate, and one bottom Te atom. The measured height differences between the top and intermediate Te atoms and between the top and bottom Te atoms are $d_{\mathrm{Top-Int}} = 9$ pm and $d_{\mathrm{Top-Bot}} = 17$ pm, respectively. These values are comparable to the picometer-scale vertical resolution limit of low-temperature nc-AFM measurements and are in reasonable agreement with the DFT-relaxed structure, which predicts $d_{\mathrm{Top-Int}} = 8$ pm and $d_{\mathrm{Top-Bot}} = 36$ pm. Additionally, the experimentally measured distance between two top Te atoms arranged in a square pattern is $a_{\mathrm{square,AFM}} = b_{\mathrm{square,AFM}} = 0.69$ nm.This value agrees well with the periodicity extracted from the STM measurements (Fig.~\ref{Fig.1}(d), left panel), which shows a constant-current STM topography acquired at negative sample bias over the same scan area. This agreement further highlights the complementarity of the two techniques.

Although the tip–sample interaction force in nc-AFM can depend on both the local electronic environment and atomic species, the topmost surface layer of TaCo$_2$Te$_2$ is composed exclusively of Te atoms, while the Co and Ta atoms are located deeper within the monolayer and therefore contribute only weakly to the measured contrast (see Fig.~\ref{Fig.1}(a)). Consequently, the relative vertical offsets observed in the line profiles are expected to predominantly reflect differences in atomic height. This strong correspondence supports our structural interpretation and provides a reliable foundation for the subsequent investigation of the electronic properties responsible for the square symmetry observed in the surface electron density.

\subsection*{Surface electronics properties}

\begin{figure}[h!]

\begin{center}
\includegraphics[width=\columnwidth]{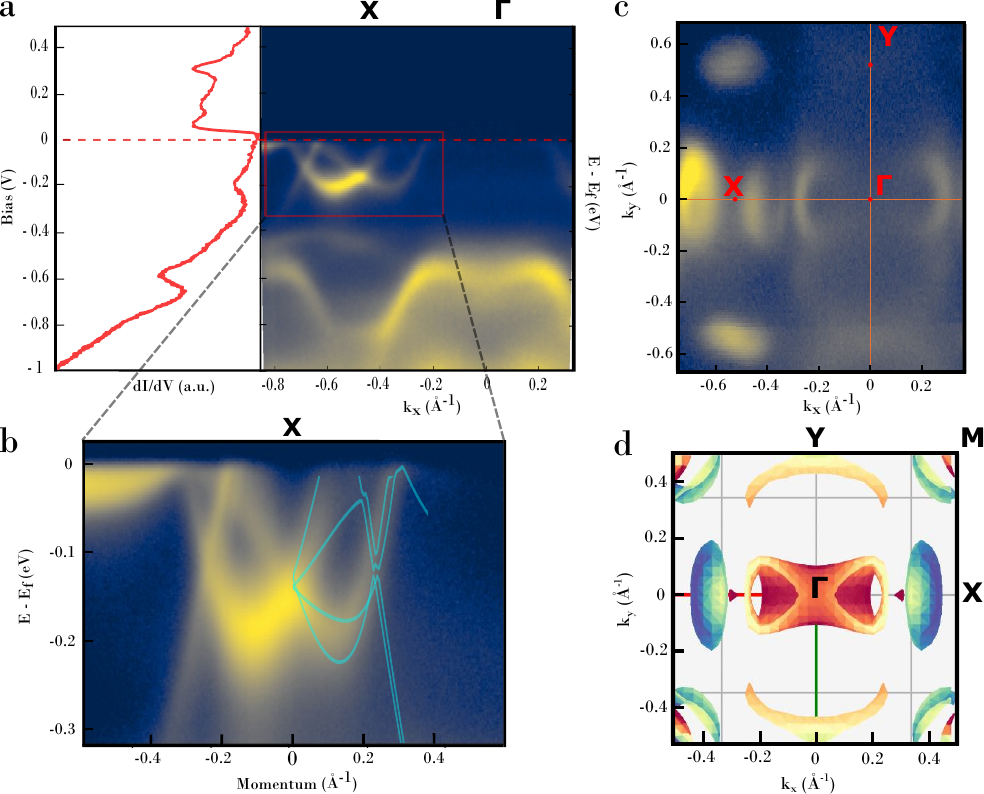}
\caption{(a) Left : Differential conductance (dI/dV) spectrum acquired by scanning tunneling spectroscopy on a pristine region of TaCo$_2$Te$_2$. Right : ARPES Band structure of TaCo$_2$Te$_2$ measured at T = 16K, along the $\Gamma - $ X direction. (b) Same ARPES band structure of TaCo$_2$Te$_2$ in a window energy of 0.0 and -0.3 eV. A DFT calculate band structure of the TaCo$_2$Te$_2$ is displayed overlying the ARPES image. (c) Fermi Surface mapping of TaCo$_2$Te$_2$ measured at T = 16K in ARPES with a photon energy of h$\nu = 25$ eV. (d) DFT simulated Fermi surface of TaCo$_2$Te$_2$, with high symmetry points indicated for clarity.}
\label{Fig.2}
\end{center}
\end{figure}

The STS spectrum acquired on a pristine region of TaCo$_2$Te$_2$ is presented in Fig.~\ref{Fig.2}(a), left panel. Several distinctive features are observed, particularly near the Fermi level. Most notably, the spectrum exhibits a depletion near Fermi energy approximately 100 mV wide, within which no pronounced low-energy excitations are detected. In addition, two asymmetric peaks appear near the Fermi level at approximately –200 mV and +60 mV, for negative and positive sample bias, respectively. Further spectral features are also visible at higher energies, including structures around –600 mV and $\approx$\ 370 mV.

STS spectra acquired under an external magnetic field of 8 T, shown in Fig. SI-2(a-b), reveal no discernible differences compared to the zero-field measurements within the experimental resolution. The absence of a measurable magnetic-field dependence does not exclude the possibility of a canted antiferromagnetic ground state. In fact, such robustness against magnetic field is consistent with the magnetization measurements reported in \cite{singha_taco2te2_2022}, where no magnetic saturation was observed up to ±8 T. On the other hand, a nonmagnetic ground state for TaCo$_2$Te$_2$ has also been proposed previously \cite{rong_realization_2023}. To further clarify the nature of the ground state, we performed DFT calculations including a Hubbard $U = 2$ eV applied to the Co 3d orbitals. The calculations indicate that the distorted TaCo$_2$Te$_2$ structure remains nonmagnetic. In contrast, when the structure is artificially constrained to an undistorted phase, analogous to TaNi$_2$Te$_2$ (see Fig. SI-3(a-d)), the system relaxes into a magnetic ground state with a magnetic moment of approximately $1\ \mu_B$ per Co atom. However, this magnetic configuration increases the total energy of the system by more than 100 meV, indicating that the distorted nonmagnetic phase is energetically favored.

The band structure of TaCo$_2$Te$2$ along the $\Gamma-X$ direction measured with ARPES is shown in the right panel of Fig.~\ref{Fig.2}(a), extending from the Fermi level down to approximately 1 eV binding energy. A double Dirac-cone-like feature is observed at the $X$ point, reaching its minimum near -0.2 eV, together with several broader dispersive bands exhibiting maxima around -0.6 eV. These observations are consistent with previous reports \cite{rong_realization_2023,pradhan_magnetic_2026}. In addition, weaker cone-like features can be identified around $ k_x \approx 0.65\ \mathrm{\AA}^{-1}$ and, less distinctly, near $ k_x \approx 0.35\ \mathrm{\AA}^{-1}$. The momentum scale observed by ARPES is consistent with the square lattice periodicity measured by STM. For a square superstructure with lattice parameter $a_{\mathrm{square}} = 0.67$ nm, the corresponding reciprocal lattice vector is $G \approx 0.98\ \mathrm{\AA}^{-1}$, yielding a Brillouin-zone boundary at $G/2 \approx 0.49\ \mathrm{\AA}^{-1}$. This agreement confirms that the low-energy electronic structure probed by ARPES is governed by the same square superstructure resolved in real space by STM.

The DFT-calculated band structure for the distorted phase of TaCo$_2$Te$_2$ is shown in Fig.~\ref{Fig.2}(b), focusing on the energy range where the Dirac-like features are observed experimentally. Overall, good agreement between the ARPES spectra and the DFT calculations is obtained (as well as in the $M-Y$ direction, shown in Fig. SI-4), particularly for the bottom of the double Dirac cone located near -200 meV, which coincides with the prominent peak observed in the STS measurements. The calculations further reveal that the Dirac-like dispersion at the $X$ point originates from two closely spaced hybridized bands forming a double-cone structure. Similar cone-like dispersions are also reproduced for the additional features around $ k_x \approx 0.35\ \mathrm{\AA}^{-1}$. Moreover, the calculations indicate the presence of small hybridization gaps, on the order of a few meV, within these cone-like states.

Interestingly, comparison between the undistorted and distorted structures (Fig. SI-6d) shows that the Dirac point at the $X$ point remains remarkably robust against the lattice distortion. We emphasize this feature in the figure, as the degeneracy is preserved by the symmetry of the distortion and therefore survives despite the structural reconstruction. While the distortion modifies portions of the electronic structure, the Dirac degeneracy itself remains intact, indicating that the underlying symmetry protection is preserved. This robustness demonstrates that the lattice reconstruction does not destroy the Dirac character of the low-energy states at $X$. These results highlight the important role of the lattice distortion in shaping the electronic structure of TaCo$_2$Te$_2$ while preserving the symmetry-protected Dirac crossing.

ARPES additionally provides direct access to the in-plane electronic momentum distribution at the Fermi level. The measured Fermi surface is displayed in Fig.~\ref{Fig.2}(c), together with the corresponding DFT simulation in Fig.~\ref{Fig.2}(d). A pronounced electronic anisotropy is observed: along the $\Gamma-X$ direction, several electron pockets are present, whereas along the $\Gamma-Y$ direction the spectral weight near the Fermi level is strongly suppressed. Comparison with the real-space structural distortion reveals an important distinction between the atomic lattice and the electronic structure probed by ARPES. Since ARPES probes momentum-resolved electronic states rather than the real-space atomic corrugation resolved by STM and nc-AFM, no direct hexagonal lattice signature is observed in the measured Fermi surface. Instead, the electronic structure near the Fermi level exhibits an approximately square symmetry, with the M point located inside an electron pocket along the ($\pi,\pi$) direction.

\begin{figure*}[h!]
\begin{center}
\includegraphics[width=1\columnwidth]{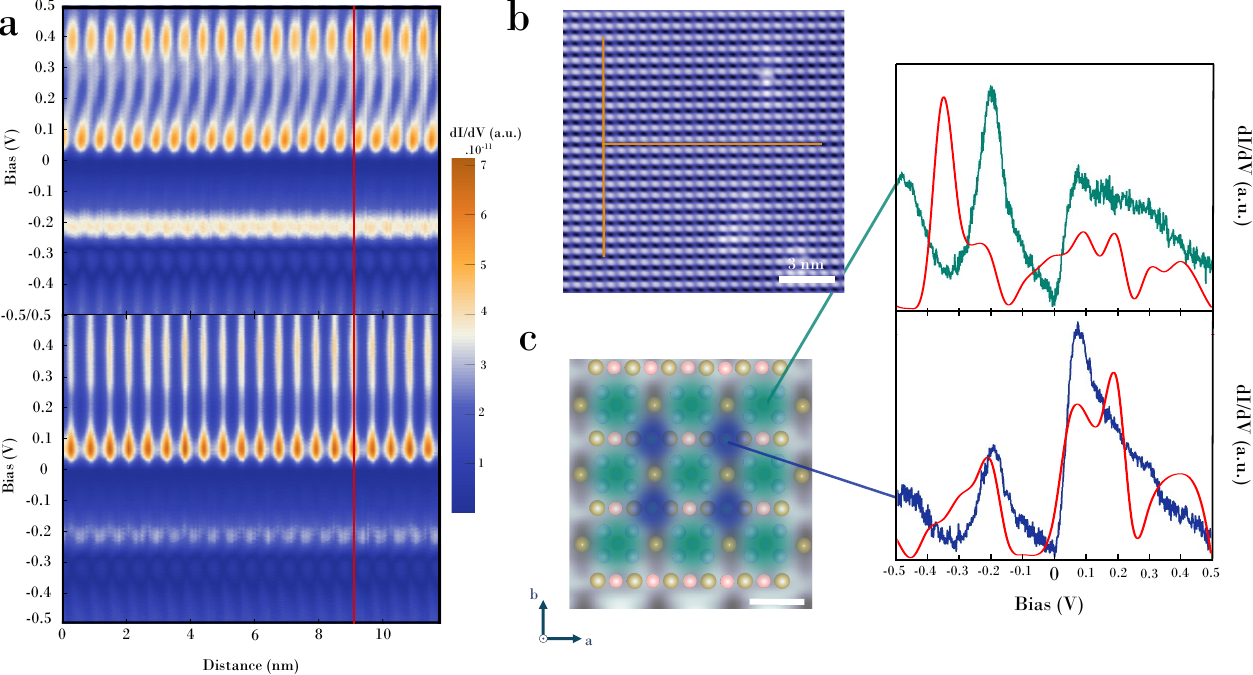}
\caption{(a) Two line STS plot as a function of bias and distance over the same length. The upper (resp. lower) trace was taken along the horizontal (resp. vertical) line located between atomic rows, as displayed in (b). The red line is a visual guide marking a fixed position. (b) Constant current image of TaCo$_2$Te$_2$ surface (-100mV, 5nA, $8\times8$ nm$^2$). (c) Constant current STM image (-100mV, 5nA, $1.7\times1.7$ nm$^2$) with superimposed atomic lattice. Two different regions are highlighted with green and blue auras and connected to their corresponding STS spectra. Right panel of (c) :  Differential conductance spectrum of TaCo$_2$Te$_2$ acquired on the pristine lattice at two distinct positions highlighted by the green and blue colors in (c). Red curves are simulated LDOS with gaussian smearing factor of 50 meV, taken 3 Å above the same positions considered in the experiment.}
\label{Fig.3}
\end{center}
\end{figure*}

Another notable feature of the STS data is the periodic modulation of the local density of states (LDOS) following the atomic lattice as shown plots of differential conductance as a function of both bias voltage and position in Fig.~\ref{Fig.3}(a) and Fig.~SI-5(a,b). The STS line scans were performed both on and between the atomic rows along the square lattice main directions (Fig.~\ref{Fig.3}(b) and Fig.~SI-5(c)). Focusing on the two peaks that are flanking the Fermi energy (located at –200 mV and +60 mV), the spatially resolved spectra shown in Fig.\ref{Fig.3}(c) display two distinct extrema. The peak below the Fermi level reaches its maximum intensity on the bright centers, while the peak above E$_\mathrm{F}$ is strongest in between four bright centers, i.e. midway between neighboring sites. 
The contrast rotation observed in the STM topography therefore does not originate from a simple reversal between occupied and unoccupied states, but rather from the energy-integrated LDOS dominated by specific electronic states exhibiting opposite spatial contrast at selected energies, mainly around -200 mV and 60 mV. Figure \ref{Fig.3}(c) also presents the calculated LDOS evaluated at a fixed height above the surface for the same spatial positions probed experimentally. Apart from a rigid energy shift, which is also required to align the calculated and experimental band structures, excellent agreement is obtained between the calculated LDOS and the STS spectra, including the presence and spatial evolution of the characteristic extrema.

\begin{figure}[h!]

\begin{center}
\includegraphics[width=\columnwidth]{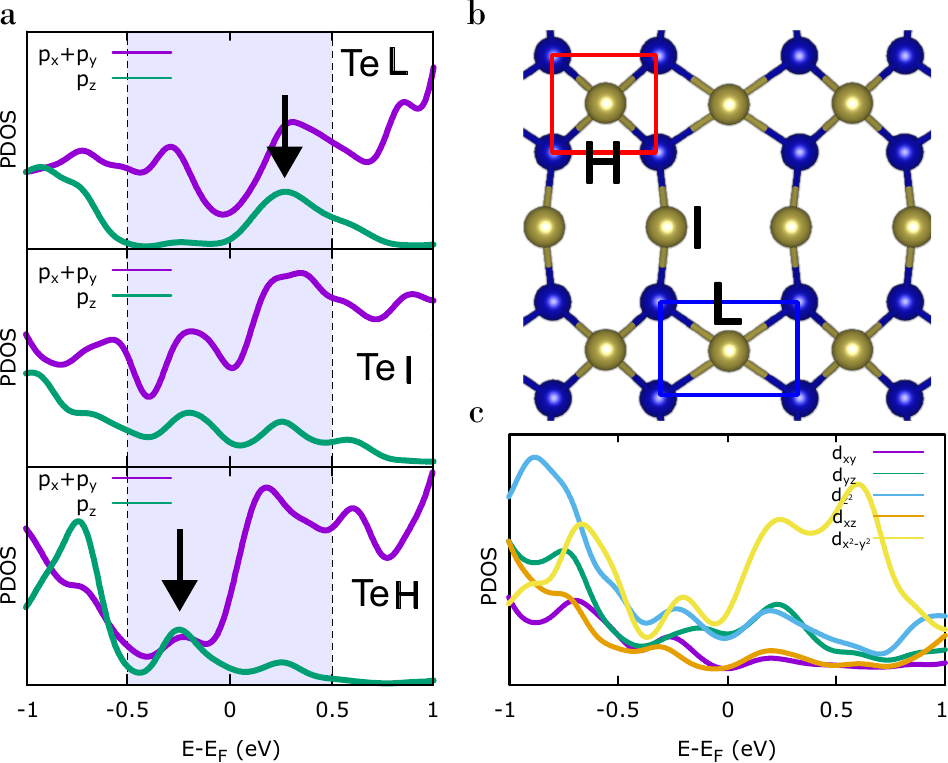}
\caption{(a) Projected Density of States (PDOS) on the \textit{p} orbitals of the low, intermediate and high height Te sublattices, respectively, as depicted in the structure shown in (b). (b) Top view of only the Te (yellow) and Co (blue) layers of TaCo$_2$Te$_2$. (c) PDOS onto the \textit{d} orbitals of the Co atoms of TaCo$_2$Te$_2$. The black arrows in (a) indicate the relevant contributions to the PDOS in the energy window between -0.5 and 0.5 eV.}
\label{Fig.extra}
\end{center}
\end{figure}

The microscopic origin of the observed LDOS modulation can be better understood with the projected density of states (PDOS) of selected surface atoms, which is shown in Figure.\ref{Fig.extra}(a-c). The calculations reveal that the Te atoms are distinguished not only by their relative vertical position, but also by their hybridization with the underlying Co atoms in TaCo$_2$Te$_2$. The higher (H) Te atoms are located above square Co arrangements (highlighted in red in Figure.\ref{Fig.extra}(b)), and the Te-Co bond length is 2.51 Å, whereas the lower (L) Te atoms are located above rectangular Co configurations (highlighted in blue in Figure.\ref{Fig.extra}(b)), with a slightly larger Te–Co bond length of 2.59 Å. This structural asymmetry breaks the equivalence between the fourfold-coordinated Te atoms. Although the bond elongation for the L Te atoms is relatively small, it is sufficient to stabilize the distorted geometry and produce distinct electronic signatures for the H and L sites. The PDOS analysis further shows that the $p_z$ orbitals of the H Te atoms contribute predominantly below the Fermi level, whereas the $p_z$ contribution from the L Te atoms becomes stronger at positive bias. This inversion of spectral weight naturally explains the bias-dependent contrast observed in the STM images shown in Fig.~\ref{Fig.1}(c). For negative sample bias near the Fermi level, the H Te atoms dominate the tunneling signal, while at positive bias the L Te atoms become more prominent. Conversely, the undistorted phase of TaCo$_2$Te$_2$ (analogous to the structure of TaNi$_2$Te$_2$) does not distinguish between H and L Te atoms, since the square and rectangular Co environments become equivalent in the absence of the distortion (see Fig.~SI-6). The intermediate (I) Te atoms differ from both H and L sites, as they are coordinated to only two Co atoms with a bond length of 2.53 Å. Their $p_z$ projected DOS remains comparatively similar for positive and negative energies around the Fermi level. Interestingly, both the distorted and undistorted phases of TaCo$_2$Te$_2$ retain the I-type Te atoms, suggesting that the primary electronic reconstruction is associated with the symmetry breaking between the H and L sites. Finally, all Co atoms in the distorted phase exhibit nearly identical PDOS, characterized by strong hybridization among the Co d orbitals. Most of the d-orbital states are fully occupied, with the exception of $d_{x^2-y^2}$ orbital, which retains a substantial spectral weight above the Fermi level. This partial occupation is closely related to the emergence of magnetization in the undistorted phase of TaCo$_2$Te$_2$, as discussed in more detail in the Supplementary Information (see Fig.~SI-6).

\subsection*{Discussion}

The combined experimental observations in TaCo$_2$Te$_2$ point toward a strong interplay between lattice reconstruction and electronic hybridization involving Co and Te states. STM topography, even at low bias and high tunneling current, does not directly resolve the hexagonal atomic lattice of the surface Te atoms, but instead reveals a pronounced square-like electronic pattern. This immediately suggests that the low-energy electronic structure is governed by an emergent reconstructed electronic potential rather than by the surface atomic lattice alone. Consistently, constant-$\Delta f$ nc-AFM measurements reveal a distortion of the top Te layer that partially follows the square symmetry of the underlying Co arrangement.

The lattice reconstruction modifies the local electronic environment and strongly influences the low-energy electronic structure near the Fermi level. In agreement with this picture, spatially resolved STS measurements reveal a modulation of the local density of states with a square-like symmetry, with spectral weight maxima located above positions associated with the underlying Co lattice and between neighboring atomic sites. These observations suggest that the low-energy electronic structure is dominated by anisotropic hybridized orbitals, giving rise to an emergent reduced symmetry despite the intrinsically hexagonal lattice geometry.

Further support for this interpretation is provided by the square-like anisotropy observed in the ARPES Fermi surface, together with the factor-of-two relation between the characteristic momentum scale observed in ARPES and the real-space periodicity resolved by STM. Combined with the DFT calculations, these results indicate that strong hybridization between Co (d) and Te (p) orbitals plays a central role in the reconstructed electronic structure near the Fermi level. The calculations additionally show that the undistorted structure neither reproduces the experimentally observed STM contrast nor stabilizes the same electronic ground state, instead favoring magnetic instabilities. The suppression of magnetic tendencies in the distorted phase therefore suggests that the lattice reconstruction and the associated electronic hybridization act cooperatively to stabilize the observed ground state by lowering the total electronic energy relative to competing magnetic configurations.

\subsection*{Conclusion}
In summary, our combined STM, nc-AFM, STS, ARPES, and DFT investigations reveal that the electronic structure of TaCo$_2$Te$_2$ is strongly governed by a surface lattice reconstruction accompanied by pronounced hybridization between Co (d) and Te (p) orbitals. Although the surface Te atoms retain an overall hexagonal arrangement, both the real-space electronic structure and the momentum-resolved electronic states exhibit an emergent square-like symmetry.

The spatial modulation of the LDOS observed in STS, the square-like anisotropy of the ARPES Fermi surface, and the agreement with DFT calculations collectively indicate that the low-energy states are dominated by hybridized Co–Te orbitals rather than by the atomic lattice geometry alone. Furthermore, the distorted phase suppresses the magnetic instability predicted for the undistorted structure, suggesting that the lattice reconstruction and the associated orbital hybridization cooperatively stabilize a nonmagnetic ground state by lowering the total electronic energy.

More broadly, our results demonstrate how lattice reconstruction and orbital hybridization can strongly modify the electronic and magnetic ground states of layered quantum materials. The sensitivity of TaCo$_2$Te$_2$ to structural distortion suggests that external tuning parameters such as strain, reduced dimensionality, or interface engineering may provide viable routes to manipulate the balance between hybridization and magnetism, potentially enabling the controlled emergence of magnetic phases in this system.

\section*{Methods}

\emph{Scanning tunneling microscopy (STM) and spectroscopy (STS) measurements:} 
Single crystals of TaCo$_2$Te$_2$ were obtained from 2D Semiconductors Inc. The samples were cleaved \textit{in situ} under ultrahigh vacuum (UHV) conditions at room temperature using adhesive Scotch tape, owing to the easily cleavable nature of the sample. The freshly cleaved crystals were subsequently transferred into a low-temperature scanning tunneling microscope (CreaTec LT-STM) connected to the same UHV system. STM experiments were performed at temperatures of $T = 4$ K and at a pressure lower than $5.10^{-11}$ mbar.
Topographic images were acquired in constant-current mode unless otherwise specified. Differential conductance (d$I$/d$V$) spectra were measured using a standard lock-in technique in open-feedback-loop conditions while sweeping the sample bias. For large-energy-range spectra, a bias modulation of 15–20 mV (peak-to-peak) at a frequency of 748 Hz was applied. Magnetic fields up to 9 T perpendicular to the sample plane could be applied during the measurements. STM data were processed and analyzed using WSxM software and custom-developed MATLAB routines.

\emph{Atomic Force Microscopy (AFM)} : 
The CreaTec LT-STM systems described above were additionally equipped with qPlus sensors for non-contact atomic force microscopy (nc-AFM) measurements. The experiments were performed using a qPlus sensor with a resonance frequency $f_0$ = 29.19 kHz, a quality factor $Q \approx 10^{5} - 10^{6}$ and a spring constant of $k = 1.8\ kN.m^{-1}$. nc-AFM images were acquired either in constant-height mode by recording the frequency shift of the qPlus sensor during scanning, or in constant-frequency-shift mode by maintaining a fixed frequency shift while mapping the surface topography. AFM data were analyzed using WSxM software and custom-developed MATLAB routines.

\emph{Angle resolved photoemission spectroscopy measurements:} 
ARPES measurements were performed in Synchrotron SOLEIL at the CASSIOPEE beamline with a SCienta R4000 hemispherical analyzer and acquired at a temperature of 16 K. A linear polarized light has been used with a photon energy of $ h\nu \ = \ 25\ eV $. The single crystal of TaCo$_2$Te$_2$ were cleaved in situ with the top-post technique at a temperature of 16 K and a pressure of $5.10^{-11}$ mbar. The ARPES data have been processed and analyzed using NAVARP tool for python.

\emph{DFT calculations:} 
Calculations were performed with the DFT methodology as implemented in the periodic plane-wave-basis VASP code \cite{PhysRevB.54.11169,KRESSE199615}. Atomic positions and lattice parameters were obtained by fully relaxing all structures using the spin-polarized Perdew–Burke–Ernzehof (PBE) functional \cite{PhysRevLett.77.3865} including Grimme's semiempirical DFT-D3 scheme for dispersion correction \cite{10.1063/1.3382344}, specially important to describe the van der Waals interactions between the sextuple layers. The interactions between electrons and ions were described by PAW pseudopotentials, and an energy cut-off of 500 eV was used to expand the wavefunctions. For structural relaxations, a $8\times8\times3$ k-grid was used for the bulk structures, while $8\times8\times1$ was used for the monolayers. Twice denser k-grids were used for band structure and density of states calculations. The convergence criterion of self-consistent field computation was set to $10^{-5}$ eV, and the threshold for the largest force acting on the atoms was set to less than 0.01 eV Å$^{-1}$. For the monolayer, a vacuum layer of 15 Å was added to avoid mirror interactions between periodic images. At first, spin polarization was considered in all calculations, where an initial magnetization of 3$\mu$B was considered per Co atom and 0 otherwise. Once it was established that the ground state is non magnetic removed the spin polarization freedom from the calculations. DFT+U calculations were realized based on the approach introduced by Dudarev \textit{et al} \cite{PhysRevB.57.1505}. Scanning tunneling microscopy and local density of states were simulated within the Tersoff-Hamann approximation \cite{PhysRevB.31.805} implemented in the \textit{critic2} code \cite{OTERODELAROZA20141007,OTERODELAROZA2009157}. The Fermi surface was calculated using a dense k-grid of $19\times19\times7$ and treated with a script adapted from \url{https://github.com/QijingZheng/VASP_FermiSurface}. Images of the structures were created using VESTA \cite{https://doi.org/10.1107/S0021889811038970}.

\section*{Supporting Information}
Supporting Information is available from the Wiley Online Library or from the author.

\section*{Acknowledgements}
\textbf{Funding:} This research made use of Nanoscience Center (NSC) at the University of  Jyväskyl\"a (JYU) facilities and was supported by the European Research Council (ERC-2021-StG
No. 101039500 “Tailoring Quantum Matter on the
Flatland”), Research Council of Finland (Academy Project Nos. 3370910 and 369367)
and the Research Council of Finland through the Finnish Quantum Flagship project (Project Nos. 359240).  We acknowledge Finnish Centre of Excellence in Quantum Materials (QMAT) and CSC–IT Center for Science, Finland, for computational resources and the Aalto Science-IT project. We acknowledge SOLEIL synchrotron for provision of synchrotron radiation facilities under the proposal number No. 20241153.

\section*{Data availability}
	All the data supporting the findings are available from the corresponding authors upon request.

\section*{Conflict of Interest}
The authors declare no conflict of interest.

%\section*{Author Contributions}

\bibliographystyle{angew}
\bibliography{ArticleTaCo2Te2}

\section*{Table of Contents}

\begin{figure}[h!]
    \centering
    \includegraphics[width=0.6\textwidth]{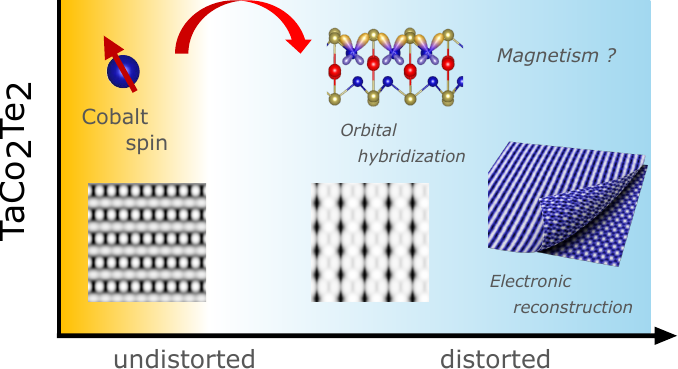}
    
    \label{TOC}
\end{figure}

Using a combination of low-temperature STM, AFM, ARPES, and DFT, we uncover a coupled electronic and structural reconstruction in TaCo$_2$Te$_2$. The reconstruction is shown to originate from an orbital-selective distortion, providing a plausible explanation for the absence of magnetism in this material.

\end{document}